\documentclass[english, 12pt]{article}
\usepackage[T1]{fontenc}
\usepackage[utf8]{inputenc}
\usepackage[english]{babel}
\usepackage{lmodern}
\usepackage[top=1in, bottom=1in, left=1in, right=1in]{geometry}
\usepackage[pdftex]{graphicx}
\usepackage{verbatim}
\usepackage[width=0.8\textwidth]{caption}
\usepackage{amsmath,amssymb,amsopn,dsfont,mathrsfs, bm}
\usepackage{float,placeins,flafter,longtable,array,booktabs}
\usepackage{color}
\usepackage{natbib}
\usepackage{xr}
\usepackage[hidelinks]{hyperref}
\makeatletter
\newtheorem{thm}{Theorem}

\newtheorem{hyp}{Assumption}

\newcommand{\R}{\mathbb R}

\newcommand{\Deriv}[2]{\frac{\partial #1}{\partial #2}}

\newcommand{\sgn}{\text{sgn}}

\newcommand{\convNor}[1]{\stackrel{d}{\longrightarrow} \mathcal{N}\left(0,#1\right)}

\renewcommand{\section}{\@startsection{section}{2}{0mm}{-1.5\baselineskip}{1\baselineskip}{\normalfont\large\bfseries}}
\renewcommand{\subsection}{\@startsection{subsection}{2}{0mm}{-1.2\baselineskip}{1\baselineskip}{\normalfont\normalsize\bfseries}}
\renewcommand{\subsubsection}{\@startsection{subsubsection}{3}{0mm}{-0.8\baselineskip}{0.4\baselineskip}{\normalfont\normalsize\itshape}}
\allowdisplaybreaks

\title{Difference-in-Differences Estimators with Continuous Treatments and no Stayers}
\author{Cl\'{e}ment de Chaisemartin, Xavier D'Haultf\oe{}uille and Gonzalo Vazquez-Bare\thanks{Chaisemartin: Sciences Po Paris, clement.dechaisemartin@sciencespo.fr. D'Haultf{\oe}uille: CREST-ENSAE, xavier.dhaultfoeuille@ensae.fr. Vazquez-Bare: University of California, Santa Barbara, gvazquez@econ.ucsb.edu.}}
\date{}

\begin{document}

\maketitle

Many treatments or policy interventions are continuous in nature. Examples include prices, taxes or temperatures. Empirical researchers have usually relied on two-way fixed effect regressions to estimate treatment effects in such cases, see e.g. \cite{deschenes2012economic}. However, such estimators are not robust to heterogeneous treatment effects in general \citep{de2020two}; they also rely on the linearity of treatment effects. We propose estimators for continuous treatments that do not impose those restrictions, and that can be used when there are no stayers: the treatment of all units changes from one period to the next. This is for instance the case when the treatment is precipitations or temperatures: for instance, temperatures of all US counties change, if ever so slightly, between two consecutive years. We start by extending the nonparametric results of \cite{chaisem_continuous} to cases without stayers. We also present a parametric estimator, and use it to revisit \cite{deschenes2012economic}.


\section{Set-up, assumptions and parameter of interest}

A representative unit is drawn from an infinite super population, and observed at two time periods. All expectations below are taken with respect to the distribution of variables in the super population. We are interested in the effect of a continuous and scalar treatment variable on that unit's outcome. Let $D_t$  denote the unit's treatment at period $t\in\{1,2\}$ and let $\mathcal{D}_t$ denote its support; let also $\mathcal{D}$ denote the support of $(D_1,D_2)$. For any $(d_1,d_2)\in \mathcal{D}$, let $Y_t(d_1,d_2)$  denote the unit's potential outcome at $t$ with treatment $d$, and let $Y_t$ denote their observed outcomes: $Y_t=Y_t(D_1,D_2)$. Finally, for any random variables $(X_t)_{t=1,2}$, let $\Delta X=X_2-X_1$. We impose the following assumptions:

\begin{hyp}\label{hyp:no_antic_static} (Static model)
For all $t\in\{1,2\}$ and $(d_1,d_2)\in \mathcal{D}$, $Y_t(d_1,d_2)$ only depends on $d_t$; we denote it by $Y_t(d_t)$.
\end{hyp}

\begin{hyp}\label{hyp:parallel_trends} (Parallel trends)
$\forall d\in \mathcal{D}_1$,  $E(\Delta Y(d)|D_1=d,D_2)=E(\Delta Y(d)|D_1=d)$.
\end{hyp}

\begin{hyp}\label{hyp:regularity} (Bounded treatment, bounded-lipschitz potential outcomes)
\begin{enumerate}
\item $\mathcal{D}_1$ and $\mathcal{D}_2$ are bounded subsets of $\mathbb{R}$.
\item $\exists \overline{Y}\ge 0$: $\sup_{(d_1,d_2)\in \mathcal{D}}E[\overline{Y}|D_1=d_1, D_2=d_2]<\infty$, and $\forall (t,d,d')\in \{1,2\}\times \mathcal{D}_t^2$, $|Y_t(d)-Y_t(d')|\leq \overline{Y}|d-d'|$.
\end{enumerate}
\end{hyp}
Assumptions \ref{hyp:parallel_trends}-\ref{hyp:regularity} are also imposed by \cite{chaisem_continuous}, and are discussed therein.
\begin{hyp}\label{hyp:nostayers} (No stayers but quasi-stayers)
$P(\Delta D= 0)=0$, $P(|\Delta D| \le \eta)>0 \;\forall\eta>0$.
\end{hyp}
First, Assumption \ref{hyp:nostayers} states that there are no ``stayers'', namely units for which $D_1=D_2$. This is in contrast with \cite{chaisem_continuous}, who assume throughout that there are stayers. Second, Assumption \ref{hyp:nostayers} states that there are ``quasi-stayers'', namely units whose treatment change may be infinitesimally small. This assumption is realistic when the treatment is, say, temperatures: some counties may have very similar temperatures from one year to the next, though no county has exactly the same temperatures.

\medskip
Hereafter, we focus on the following effect:
\begin{align}
	\theta_0 = &E\left(\frac{|\Delta D|}{E(|\Delta D|} \times \frac{Y_2(D_2)-Y_2(D_1)}{D_2-D_1}\right) \label{eq:for_AME} \\[2mm]
=& \frac{E\left(\text{sgn}(\Delta D) (Y_2(D_2)-Y_2(D_1))\right)}{E(|\Delta D|)}. \notag
\end{align}
$\theta_0$ is a weighted average of the slopes of units' potential-outcome functions, from their period-one to their period-two treatment, the so-called WAOSS in \cite{chaisem_continuous}. It follows from the mean-value theorem that it may be seen as a weighted average marginal effect.

\section{Nonparametric identification and estimation} 
\label{sec:identification}

\begin{thm}\label{thm:main_nostayers}
If Assumptions \ref{hyp:no_antic_static}-\ref{hyp:nostayers} hold, $$\theta_0 = [E\left(S\Delta Y\right) - \zeta_0]/E[|\Delta D|],$$ with $S:=\sgn(\Delta D)$ and
$$\zeta_0:= E\left[S\lim_{\eta\downarrow 0} E(\Delta Y | D_1, |D_2-D_1|\le \eta)\right].$$
\end{thm}
Theorem \ref{thm:main_nostayers} shows that without stayers, $\theta_0$ is identified by the limit (as $\eta\downarrow 0$) of a difference-in-difference comparing the $\Delta Y$ of all units and of quasi-stayers.

\medskip
We now discuss estimation of $\theta_0$. Only the estimation of $\zeta_0$ raises difficulties. We show in the proof of Theorem \ref{thm:main_nostayers} that under our assumptions, $g(d_1,\delta):= E[\Delta Y|D_1=d_1,\Delta D=\delta]$ is well-defined and continuous at $(d_1,0)$, for any $d_1\in \mathcal{D}_1$. Hence, $\zeta_0$ satisfies $\zeta_0=E\left[Sg(D_1,0)\right]$. This formulation links our problem to the estimation of nonparametric additive models. To see this, suppose that the variables $(W,X)\in \R\times \R^k$ satisfy $h(x):=E[W|X=x]=\sum_{j=1}^k h_j(x_j)$ for some unknown functions $(h_j)_{j=1,...,k}$. Then, under the normalization $E[h_j(X_j)]=0$ for $j<k$, we can identify and estimate $h_k$ by remarking that
\begin{equation}
h_k(x_k) = E[h(X_1,...,X_{k-1},x_k)].
	\label{eq:marg_integ}
\end{equation}
We can then estimate $h_k(x_k)$ by first estimating $h$ with any usual nonparametric estimator, and second plugging it in the sample counterpart of the expectation in \eqref{eq:marg_integ}. As \cite{linton1995} and \cite{kong2010uniform} show, the corresponding estimator is, under regularity conditions,  asymptotically normal and converges at the standard univariate nonparametric rate (namely, $n^{2/5}$, with $n$ the sample size). This rate is also the optimal convergence rate for this problem \citep{stone1985additive}. Up to minor changes (in $\zeta_0$, $g$ plays the role of $h$ in \eqref{eq:marg_integ} and $\zeta_0$ also includes $S$), our parameter $\zeta_0$ can be obtained in the same way as $h_k(x_k)$, so we can also obtain an  asymptotically normal estimator converging at the $n^{2/5}$ rate.

\medskip
This contrasts with the standard ($n^{1/2}$) rate obtained for the estimators of the WAOSS in the presence of stayers, as shown by \cite{chaisem_continuous}. To understand the difference, note that with stayers, the proportion of units used as controls to reconstruct switchers' counterfactual outcome evolution remains positive as $n\to \infty$. On the other hand, it tends to zero here, since we need to consider quasi-stayers, with $\eta\to 0$ as $n\to \infty$ to avoid any bias. This results in a lower rate of convergence.

\medskip
Finally, in applications with no stayers, it is more difficult to propose placebo estimators of the parallel trends assumption. When a third period of data, period zero, is available, a placebo mimics the actual estimator, replacing $\Delta Y$ by units' period-zero-to-one outcome evolution. However, as units' treatments may have changed from period zero to one, one would need to restrict the sample to period-zero-to-one quasi-stayers, to avoid that the placebo differs from zero due to the treatment's effect. Thus, the placebo would compare the period-zero-to-one outcome evolution of period-one-to-two switchers and quasi-stayers, restricting the sample to period-zero-to-one quasi-stayers. Then, we conjecture that the number of units used as controls by the placebo may tend to zero faster than the number of units used as controls by the actual estimator, for instance if being a period-zero-to-one and a period-one-to-two quasi-stayer are independent events. Then, the placebo may converge at an even slower rate than the actual estimator.



\section{A parametric approach} 

We now consider a parametric root-$n$ consistent estimator, that avoids issues related to nonparametric estimation and inference, while still allowing for heterogeneous and nonlinear effects. Specifically, we impose that $g(d_1,\delta)=g_{\lambda_0}(d_1,\delta)$, where the family $(g_{\lambda})_{\lambda\in\R^p}$ is known (but $\lambda_0$ is not). By definition of $g$ and Assumption \ref{hyp:parallel_trends},
\begin{align*}
&g(d_1,\delta)=E[Y_2(d_1)-Y_1(d_1)|D_1=d_1] +\delta 
E\left[\frac{Y_2(d_1+\delta)-Y_2(d_1)}{\delta}\big|D_1=d_1,\Delta D=\delta\right].
\end{align*}
Thus, the parametric assumption amounts to imposing restrictions on both
$d_1 \mapsto E[Y_2(d_1)-Y_1(d_1)|D_1=d_1]$ and the average slope  $ (d_1,\delta) \mapsto  E[(Y_2(d_1+\delta)-Y_2(d_1))/$ $\delta |D_1=d_1,\Delta D=\delta]$.
For instance, if $g_{\lambda}(d_1,\delta)$ is linear, we assume that the former function is linear, and the latter is constant. Similarly, $g$ is a polynomial if both functions are polynomial. Note that we can test that $E[\Delta Y|D_1=d_1,\Delta D=\delta]=g_{\lambda_0}(d_1,\delta)$ for some $\lambda_0$ by a parametric specification test, see e.g. \cite{bierens1982consistent} or \cite{hong1995consistent}.

\medskip
We consider a simple two-step estimator based on this parametric restriction and an i.i.d. sample $(D_{1i},\Delta D_i, \Delta Y_i)_{i=1,...,n}$. In the first step, we estimate $\lambda_0$ by (linear or nonlinear) least squares or, more generally, a GMM estimator $\widehat{\lambda}$. In the second step, we estimate $\theta_0$ by
$$\widehat{\theta} = \frac{\sum_{i=1}^n S_i (\Delta Y_i - g_{\widehat{\lambda}}(D_{1i}, 0))}{\sum_{i=1}^n |\Delta D_i|}.$$
Since $\widehat{\theta}$ may be seen as a two-step GMM estimator, we obtain, under Assumptions \ref{hyp:no_antic_static}-\ref{hyp:nostayers} and standard regularity conditions on $\lambda \mapsto g_{\lambda}(d_1,\delta)$,
$$\sqrt{n}\left(\widehat{\theta} - \theta_0\right) \convNor{V(\psi)},$$
where the influence function $\psi$ satisfies
\begin{align*}
&\psi = \frac{1}{E[|\Delta D|]}\left[S\left(\Delta Y - g_{\lambda_0}(D_1,0)\right)
- E\left[S\Deriv{g}{\lambda}(D_1,0){}_{|\lambda=\lambda_0}\right] \times \xi - \theta_0 |\Delta D|\right],
\end{align*}
with $\xi$ the influence function of $\widehat{\lambda}$. We can thus simply estimate $V(\psi)$ by a plug-in estimator, using an initial estimator of $\xi$.

\section{Application} 
\label{sec:application}

We use the data from \cite{deschenes2012economic} to compute our parametric estimator. The authors use a balanced panel of 2,342 US counties in years 1987, 1992, 1997, and 2002, and consider TWFE regressions, weighted by counties' farmland acres, of annual agricultural profits in county $c$ and year $t$ on four treatment variables: growing season degree days, growing season degree days squared, precipitations, and precipitations squared. To fit in the two-periods-one-treatment case we consider, we restrict the data to years 1997 and 2002, and we focus on the growing season degree days treatment. The coefficient of that treatment in a TWFE regression estimated on years 1997 and 2002 and weighted by counties' farmland acres is equal to  -0.024 (s.e. clustered at the county level: 0.007), which is close to the corresponding TWFE coefficient keeping the four years and all treatments (-0.015, s.e. clustered at the county level: 0.005). Assuming that $$E[Y_2(d_1)-Y_1(d_1)|D_1=d_1]=\lambda_{0,1}+\lambda_{0,2} d_1$$
and
\begin{align*}
&E\left[\frac{Y_2(d_1+\delta)-Y_2(d_1)}{\delta}\middle|D_1=d_1,\Delta D=\delta\right]
=\lambda_{0,3}+\lambda_{0,4}d_1+\lambda_{0,5}\delta,
\end{align*}
we find that $\widehat{\theta}$, weighted by counties' farmland acres as well, is equal to $-0.018$ (s.e.: 0.011) 
Thus, the conclusion from the TWFE regression seems robust to allowing for some effect heterogeneity, even though the estimated effect is less significant. While arguably restrictive, our model for the conditional expectation function of slopes allows for some non-linearity and heterogeneity in the effects of temperatures on agricultural output.


\section*{Appendix: proof of theorem \ref{thm:main_nostayers}}

It suffices to show that a.s.,
\begin{equation}
\lim_{\eta\downarrow 0}E\left(\Delta Y|D_1,|\Delta D|\le \eta\right)=
E\left(Y_2(D_1)-Y_1(D_1)|D_1,D_2\right).	
	\label{eq:key_id_nostayers}
\end{equation}
Fix $\eta>0$. By Assumption \ref{hyp:nostayers}, $P(|\Delta D|\le \eta|D_1)>0$. Thus, $E\left(\Delta Y|D_1,|\Delta D|\le \eta\right)$ is well-defined. Moreover,
\begin{align}
E\left(\Delta Y|D_1,|\Delta D|\le \eta\right)=& 
 E\left(Y_2(D_2)-Y_2(D_1)|D_1,|\Delta D|\le \eta\right) \notag \\
& + E\left(Y_2(D_1)-Y_1(D_1)|D_1,|\Delta D|\le \eta\right).
\label{eq:decomp}
\end{align}
Now, by Jensen's inequality and Point 2 of Assumption \ref{hyp:regularity},
\begin{align}\label{eq:ft_neg}
\big|E\left[Y_2(D_2)-Y_2(D_1)|D_1,|\Delta D|\le \eta\right]\big|  
\le & E\left(\left|Y_2(D_2)-Y_2(D_1)\right| \; |D_1,|\Delta D|\le \eta\right) \notag  \\
\le & E\left(\overline{Y}|D_2 - D_1| \; |D_1,|\Delta D|\le \eta\right) \notag  \\
\le & \eta E\bigg[\sup_{(d_1,d_2)\in\mathcal{D}} E\left(\overline{Y}|D_1=d_1,D_2=d_2\right) | 
 D_1,|\Delta D|\le \eta\bigg] \notag  \\
\le & \overline{K}\eta
\end{align}
for some $\overline{K}<\infty$. Next, by Assumption \ref{hyp:parallel_trends},
\begin{align*}
E\left(Y_2(D_1)-Y_1(D_1)|D_1,|\Delta D|\le \eta\right)= & 
E\left(Y_2(D_1)-Y_1(D_1)|D_1\right) \\
=& E\left(Y_2(D_1)-Y_1(D_1)|D_1,D_2\right).
\end{align*}
Combined with \eqref{eq:decomp}-\eqref{eq:ft_neg}, this yields \eqref{eq:key_id_nostayers}\; $_\Box$

\bibliographystyle{aea}
\bibliography{biblio.bib}

\end{document}